# How is Your Mood When Writing Sexist tweets? Detecting the Emotion Type and Intensity of Emotion Using Natural Language Processing Techniques

Sima Sharifirad
Faculty of computer science,
Dalhousie university, Halifax,
Nova scotia, Canada
s.sharifirad@dal.ca

Stan Matwin
Faculty of computer science,
Dalhousie university, Halifax,
Nova scotia, Canada
stan@cs.dal.ca

Borna Jafarpour
Huawei Technologies, Noha's
Ark Lab, Toronto, Canada
jafarpour@gmail.com

## ABSTRACT

Online social platforms have been the battlefield of users with different emotions and attitudes toward each other in recent years. While sexism has been considered as a category of hateful speech in the literature, there is no comprehensive definition and category of sexism attracting natural language processing techniques. Categorizing sexism as either benevolent or hostile sexism is so broad that it easily ignores the other categories of sexism on social media. Sharifirad S and Matwin S 2018 proposed a well-defined category of sexism including indirect harassment, information threat, sexual harassment and physical harassment, inspired from social science for the purpose of natural language processing techniques. In this article, we take advantage of a newly released dataset in SemEval-2018 task1: Affect in tweets, to show the type of emotion and intensity of emotion in each category. We train, test and evaluate different classification methods on the SemEval-2018 dataset and choose the classifier with highest accuracy for testing on each category of sexist tweets to know the mental state and the affectual state of the user who tweets in each category. It is a nice avenue to explore because not all the tweets are directly sexist and they carry different emotions from the users. This is the first work experimenting on affect detection this in depth on sexist tweets. Based on our best knowledge they are all new contributions to the field; we are the first to demonstrate the power of such in-depth sentiment analysis on the sexist tweets.

## 1. INTRODUCTION

Social media has been the battlefield of users for many years. Their language indeed reveals their values, their perspective and their emotions. Among all types of hateful speech and abusive languages, sexist tweets have been very pervasive in social media platforms like Twitter and Facebook. Waseem et al. 2017 [13] were the first who collected hateful tweets and categorized them into being sexist, racist or neither. However, they did not provide specific definitions for each category. They presented eighteen conditions for all the tweets being sexist or racist, but they did not present a specific definition for each category. One year later, Jha and Mamidi (2017) [6] focused on just sexist tweets and proposed two categories of hostile and benevolent sexism. However, these categories were so general that they simply ignored other types of sexism happening in social media. In one step further, Sharifirad S. and Matwin S 2018 [10] proposed complimentary categories of sexist language inspired from social science work. They categorized the sexist tweets into the categories of indirect harassment, information threat, sexual harassment and physical harassment.

There has been a lot of work on sentiment classification in twitter datasets. It mainly focused on the sentiment as being positive, negative and neutral. As for the sexism, it can be implied that the sentiment of sexist tweets is more negative than positive. However, benevolent sexism, which implies a subjectively positive view towards men or women, is a clear example of sentences which carry neutral or positive sentiment. Saif Moahmmad et al. (2017) [9] released a new dataset in the recent SemEval-2018 Task 1: Affect in Tweets. The new dataset has new and useful aspects for emotion classification, intensity of emotion classification and intensity of sentiment classification (valence). It has been argued that these classes can reveal the mental state of the tweeter.

In this paper, we focused on Sharifirad S. and Matwin S, (2018) [10] categories of sexism and tried to predict the type of emotion, intensity of sentiment and intensity of emotion in each of the sexual harassment categories separately. Based on our best knowledge they are all new contributions; we are the first to demonstrate the mental state of the person who tweets sexist tweets by showing the type of emotion, the intensity of the emotion and the intensity of the sentiment. Contributions of the paper are as follows:

- We chose the best methods on the Semeval2018 dataset and test it on each sexism categories to know the emotion type of users who post these types of tweets.
- We choose the best methods on the Semeval2018 dataset and test it on each sexism categories to know the emotional intensity of users who post these types of tweets.
- We choose the best methods on the Semeval2018 dataset and test it on each sexism categories to know the sentiment of the tweets.
- We present a discussion based on the previous classification results on each category.

## 2. Background and Related Network

**Problem:** Over the past few years, sexism and sexual harassment against women has attracted considerable movements such as #byfelip, #metoo #mencallmethings. Yet, there have not been many studies to address this issue focusing on natural language processing techniques. Other related works focused on abusive detection [14] hate speech detection [12, 4], racism and sexism behavior [13, 6, 2], hostile and benevolent sexism [6] and in a more focused work on different sexist categories [10]. However, these studies were primary and they have not focused on the sentiment of the sexist tweets.

**Methods:** Waseem et al. (2017) [13] were the first who used natural language processing techniques on the categories of sexism, racism and neither. They published the tweets with their tweet Id for other researchers to use. After that, many methods have been proposed ranging from traditional methods to deep learning methods. Jha and Mamidi (2017) [6] tested support vector machine, bi-directional RNN encoder-decoder and FastText on hostile and benevolent sexist tweets. They also used SentiWordNet and subjectivity lexicon on the extracted phrases to show the polarity of the tweets. An ensemble method, including Long Short Term Memory (LSTM) based classifiers and a set of user historical features [2]. Clarke and Grieve (2017) [15] worked on Multi-Dimensional Analysis (MDA) of the racist and sexist tweets and show their difference in three dimension of being interactive, antagonistic and attitudinal. In another study, Gamback and Sikdar (2017) [1] trained Convolutional Neural Networks (CNN) on word grams and character grams on racist and sexist tweets.

Sentiment is the user inclination when facing a specific news or text. SemEval Task1: affect in tweets was one of the shared task datasets released by Mohammad and Kiritchenko (2018) [9] for the competition. They used best worst scaling as their annotation method and released datasets for detecting the emotion type (multi-label classification) and detecting intensity of emotion (multiclass classification). As for the first class, detecting the emotion, tweets can be categorized in eleven categories as anger, anticipation, disgust, fear, joy, love, optimism, pessimism, sadness, surprise and trust. As for the second class of dataset, four datasets were released for detecting the emotional intensity of users such as anger, fear, joy and sadness. These datasets have four classes of intensity from zero to three. Zero shows no inferred emotion, one represents low amount of that emotion, two shows moderate amount of emotion and three shows the highest amount of emotion for each emotion: anger, fear, joy and sadness. Table 1 shows the distribution of the data in each class in the training, validation and test set.

**Contributions**: To the best of our knowledge, the present work is the first study in depth to show the application of affect classification on different sexist categories. Considering the previous works, the tweets were classified into racist, sexist or neither of the groups. There was just one study focused solely on the sexist tweets and showing the percentage of polarity in each category. However, these studies lack further investigation on different types of sexist tweets and their emotion and intensity of emotion. We focused on the sexist dataset presented in Table 2. Since the number of data in the second category is not a lot, we performed the implementation on the three classes of #1, #3 and #4.

| Table2. The detail information of the sexist data distribution. ||
|---|---|
| Name of Categories | Number of data in each category |
| Indirect harassment(#1) | 260 |
| Information threat(#2) | 2 |
| Sexual harassment(#3) | 417 |
| Physical harassment(#4) | 123 |

| Table1.Distribution of Emotion intensity for anger, joy, sadness and fear. ||||
|---|---|---|---|
|  | Number(train) | Number(test) | Number(validation) |
| No anger can be inferred(#0) | 445 | 465 | 186 |
| Low amount of anger can be inferred(#1) | 322 | 148 | 54 |
| Moderate amount of anger can be inferred(#2) | 507 | 243 | 97 |
| High amount of anger can be inferred(#3) | 427 | 146 | 51 |
| No joy can be inferred(#0) | 548 | 194 | 55 |
| Low amount of joy can be inferred(#1) | 362 | 333 | 95 |
| Moderate amount of joy can be inferred(#2) | 346 | 360 | 89 |
| High amount of joy can be inferred(#3) | 359 | 218 | 51 |
| No sadness can be inferred(#0) | 594 | 398 | 170 |
| Low amount of sadness can be inferred(#1) | 260 | 193 | 88 |
| Moderate amount of sadness can be inferred(#2) | 364 | 255 | 87 |
| High amount of sadness can be inferred(#3) | 315 | 129 | 52 |
| No fear can be inferred(#0) | 1490 | 633 | 186 |
| Low amount of fear can be inferred(#1) | 320 | 124 | 54 |
| Moderate amount of fear can be inferred(#2) | 249 | 158 | 97 |
| High amount of fear can be inferred(#3) | 193 | 71 | 51 |

## 3. Text Preprocessing

Preprocessing of the tweets involves removal of the punctuation, hyperlinks/URLs, emoji and tags. Stop words were not removed because some stop words like "not" were very important for the sentiment of the sentence. Before training the classification models, WordNet lemmatization was applied on all the tweets. We set the maximum size of each tweet to 40 words, and padded the tweets of shorter length with zeroes. Next, tweets were converted into the vectors using Word2vec [11], Glove [17], FastText [18] all with the length 300.

## 4. Dataset

In this research, we focused mainly on two categories of dataset. The first is related to the two tasks of emotion detection type and emotional intensity. The second dataset is related to sexist dataset; it comprises of the four categories mentioned in table 2. We trained classification algorithms on all the initial tasks (detecting emotion type, intensity of sentiment) chose the one with highest accuracy and then tested it with the sexist tweets to know the category of emotion and intensity of emotion in each category. Table 3 shows the LDA result on the three sexist categories of indirect harassment, sexual harassment and physical harassment.

## 5. Methodology

For multiclass classification and multi label classification, we considered a baseline along with some traditional classification algorithms utilized for this purpose and deep learning algorithms described below:

**One-vs.-rest (OVR)** we trained and evaluated K independent binary classifiers for each class separately for our multi label and multiclass classification tasks. We considered all the samples in that class positive and the rest negative using LinearSVC in the Sklearn python package.

**Support vector Machines (SVM)** is used as a supervised model for the classification [3]. The main idea of the algorithm is to maximize the minimum distance from the hyper-plane, which separates the samples to the nearest sample. To classify the tweets, we used different word vectors and we considered labels as one hot encoding and multi hot encoding for multi class and multi label classification.

**Naive Bayes (NB)** the main idea behind naive Bayes is to maximize the posteriori (MAP). Based on the number of class known as prior probabilities, a class label is assigned to the new sample with its features, in order to maximize the posterior probability given the new sample. In fact, it computes the class conditional probabilities of the features having the available classes.

**K-Nearest Neighbor (KNN)** This algorithm is considered as a non-parametric classification algorithm. This algorithm usually calculates the Euclidean distance between the new sample and every other training example. The k smallest distance along with the most represented class are considered as the class label.

**Multilayer perceptron (MLP)** This algorithm is in the category of supervised algorithms. It is made in a way that can be changed easily for the multiclass classification task. In the case of multiclass classification, the word vectors (input) are multiplied by different weight vectors to calculate the activation function for each specific data point. The weight vector, which produces the highest activation, will be the class data that the sample belongs to. It requires multiple training iterations to learn the data.

**Long-short-term-memory (LSTM)** This algorithm is in the category of recurrent neural networks and uses internal memory to deal with different sequence of inputs. At the same time, it can capture dependencies very well.

**Convolutional Neural network (CNN)** This algorithm is one of the most important algorithms in computer vision and recently has been used in text classification tasks. They usually have several

| Table3. Representative topics in each category. | |
|---|---|
| **Categories** | **LDA result** |
| Indirect harassment | (Girl, think, like, cook), (will, girl, say, need), (blond, know, girl, can),(girls, women, time, amp),(girls, women, just, bitch),(girl, can, cook, just)(girls, like, cook)(blond, just, dumb, babe) |
| sexual harassment | (bitch, girl, s, sex), (girl, bitch, cam, hot), (porn, girl, man, good), (girl, fuck, bitch, nake), (bitch , girl, t, ass), (girl, bitch, fuck, sexi), (bitch, sex ,girl, like), (bitch, like, ass, shit), (bitch, girl, slut, enjoy) |
| physical harassment | (girl, like, year, old), (get, girl, see), (girl, bitch, two, face), (like, blond, one, now), (girl, face, beg, park), (bitch, look, slap, go), (girl, like, ladi ,model), (girl, get, look, amp), (bitch, girl, black ), (girl, want ,watch) |

layers of nonlinear activation function. Inspired by Gamback and Sikdar, (2017)[1] we used CNN for our multi label and multiclass classifications.

## 6. Evaluation

### 6.1 Experimental Set up

We experimented and trained the classifiers on emotion type detection and emotional intensity datasets. We used different representations learning for the words. For the embedding based methods, we used Word2vec [11]. Word vectors are trained from ten million English tweets from the Edinburgh Twitter Corpus [16]. The word2vec parameters were window size of two and length of 300. We also used Glove, embedding size of 300; its embedding has been pre-trained on around 2B tweets. The other embedding was FastText; these embeddings were trained on Wikipedia pages, considering the default mode, and embedding size of 300. After getting the vector for each word, we concatenated them to get a vector for each tweet. For the out-of-words vocabulary, we considered the vector of each character in the word and concatenated them to get the same length word vector. We trained the classifiers on each dataset training set, then validated with the validation set and finally reported the accuracy on the test set.

### 6.2 Experimental Results

Initially, we tried different methods on the SemEval 2018 dataset; table 4 shows the accuracy of the methods on two tasks of emotional intensity and emotion detection type. we picked the method with the highest accuracy to test it on the three sexist categories to know the emotion type and intensity of emotion in each category. Starting from emotional intensity and the first category as "anger", the highest accuracy is about 93% using FastText as the embedding vectors and CNN as the classification algorithm. The second category of emotional intensity was "fear"; the highest accuracy is about 91% with the embedding of FastText and the CNN as the learning algorithm. Coming to the third category as "joy", the highest accuracy is about 90% using FastText as the embedding vector and CNN as the classification algorithm. For our task, fine-tuning of CNN and using a high number of epochs for learning was effective. In comparison to the baselines, deep learning methods take advantage of multiple learning and when the number of samples is not significant, using big number of epochs is helpful. The next task was related to the emotion classification; this multi-label classification was not an easy task and the accuracy was not as good as the previous task. The highest performance is related to FastText as the embedding vector and the CNN as the classification method with the accuracy of 0.85%. The last task, classification of the sentiment intensity or valence, the highest accuracy relates to the FastText embedding and CNN for the choice of classification.

After choosing our best choice of classifier and word vector, we tested the algorithm on the three sexual harassment datasets to see how each category is different from the other one in terms of emotion type and intensity of emotion. The first task was related to the emotional intensity of "anger", "fear", "joy" and "sadness", which is shown in table 4. Tweets can have a range from 0, implying no anger, to 3, showing high anger. Out of 260 total tweets, about 240 tweets were categorized as showing no anger, 89 tweets showed a slight anger in the indirect harassment tweets. In terms of emotional intensity of fear, considering the same range, about 240 tweets were categorized as showing no fear

| Table 4 accuracy on methods on Emotion intensity and emotion type ||||||||
|---|---|---|---|---|---|---|---|
| | One Vs all | SVM | NB | KNN | MLP | LSTM | CNN |
| | W2v/Glove/FastText | W2v/Glove/FastText | W2v/Glove/FastText | W2v/Glove/FastText | W2v/Glove/FastText | W2v/Glove/FastText | W2v/Glove/FastText |
| **Anger** | 0.53/0.54/0.55 | 0.62/0.64/0.67 | 0.64/0.66/0.68 | 0.65/0.66/0.68 | 0.82/0.83/0.85 | 0.86/0.86/0.88 | 0.88/0.89/0.93 |
| **Fear** | 0.60/0.63/0.65 | 0.67/0.69/0/70 | 0.73/0.75/0.77 | 0.75/0.77/0.79 | 0.82/0.84/0.86 | 0.84/0.85/0.89 | 0.88/0.89/0.91 |
| **Joy** | 0.53/0.54/0.57 | 0.62/0.63/0.66 | 0.67/0.67/0.69 | 0.65/0.66/0.68 | 0.79/0.79/0.82 | 0.79/0.83/0.86 | 0.87/0.89/0.90 |
| **Sadness** | 0.66/0.75/0.67 | 0.68/0.69/0.70 | 0.72/0.74/0.76 | 0.68/0.68/0.71 | 0.78/0.79/0.82 | 0.81/0.83/0.86 | 0.87/0.88/0.93 |
| **Accuracy on Emotion type detection** | 0.33/0.35/0.41 | 0.51/0.56/0.59 | 0.53/0.55/0.57 | 0.52/0.54/0.55 | 0.73/0.76/0.77 | 0.79/0.83/0.84 | 0.80/0.84/0.85 |

| Table 5. Final results of emotion intensity on sexist categories | | |
|---|---|---|
| **Indirect Harassment (260)** | Emotion intensity of anger(0/1/2/3) | (170/89/1/0) |
| | Emotion intensity of fear(0/1/2/3) | (240/20/0/0) |
| | Emotion intensity of joy (0/1/2/3) | (12/0/128/120) |
| | Emotion intensity of sadness(0/1/2/3) | (90/140/30/0) |
| **Sexual Harassment (417)** | Emotion intensity of anger(0/1/2/3) | (0/7/30/380) |
| | Emotion intensity of fear(0/1/2/3) | (370/47/0/0) |
| | Emotion intensity of joy (0/1/2/3) | (0/0/27/390) |
| | Emotion intensity of sadness(0/1/2/3) | (0/63/183/171) |
| **Physical Harassment (123)** | Emotion intensity of anger(0/1/2/3) | (0/0/4/119) |
| | Emotion intensity of fear(0/1/2/3) | (114/9/0/0) |
| | Emotion intensity of joy (0/1/2/3) | (0/8/10/105) |
| | Emotion intensity of sadness(0/1/2/3) | (0/12/23/88) |

while small number showed slight fear. In terms of emotional intensity of joy, around 128 tweets were categorized in the moderate level of joy. This shows that the tweets in this category imply sarcastic characteristics of the users, they tweet a positive sentence but in a sarcastic way. The last category pertains to the emotional intensity of sadness. The majority of tweets are in categories showing slight sadness in the tweets.

The second task was about the emotion type detection task; we classified the tweets into eleven categories. Tweets in the indirect harassment category, mostly carry optimistic feelings, anger and joy. In terms of intensity of sentiment, this category had the highest amount of tweets as being very negative or slightly negative. It had the highest amount of anger in the tweets. It showed almost no sense of fear in the tweets about 240 of tweets. The tweets in this category also had the highest intensity of joy and moderate intensity of joy. Most tweets showed moderate amount of sadness. In terms of emotion classification, most tweets carried disgust, joy and sadness. It makes this story in the mind that users, who send sexual harassment tweets, become angry from a tweet and enjoy harassing the opposite sex by showing deep disgust towards her. The third category shown in table 5, is physical harassment. In terms of intensity of emotion, the largest number of high intensity of sentiment is in this category. About 119 tweets show high intensity of anger, no intensity of fear, high intensity of joy and high intensity of sadness. It has the high emotion of anger, disgust and sadness. It seems there is a lot of similarity in terms of the type of emotion and intensity of emotions in these two categories.

## 7. Discussion

Overall, FastText and CNN have the best performances on the dataset. Interestingly enough, the emotion type and emotional intensity have a lot to say in each sexual harassment categories. Starting from indirect harassment, based on the original description of this type of harassment, they are not directly violent. However, they indirectly show a kind of superiority of the men over women. Based on the results, in terms of emotional intensity, in the first place they show no level of anger, fear or sadness except for joy. Intensity of joy is high in indirect harassment. In the sexual harassment category, the emotional intensity of anger, joy and sadness is high but not fear. It shows that users who send sexual harassment tweets are usually angry or sad or enjoy writing these types of tweets. However, there is no high intensity of fear. It presents that users have no fear to tweet these kinds of tweets. In the physical harassment category, there is high intensity of anger, joy and sadness but low intensity of fear. The similarity of the results in the two categories of sexual harassment and physical harassment is expected since they share many words and semantics.

The other discussion is related to the emotion type. Indirect harassment has the biggest number of tweets categorized in joy, then surprise and anticipation. There is no surprise that joy has the highest number of tweets and it is in line with the previous results. Indirect harassment contains other complimentary definitions such as when males expect women to behave in a certain way or they show their surprise in a sarcastic way in the tweets. For example, the tweet, "she plays as good as a boy", shows the surprise of the user when noticing a female plays well or the tweet, "she should come back to kitchen", shows the anticipation of male users about females. The second category, sexual harassment, has the highest number of tweets categorized as disgust, sadness and anger. These results are in line with the previous results. In addition, disgust in this category is complimentary to the other emotions. In the physical harassment category, the highest number of tweets are categorized as anger, disgust and sadness. This result

| Table 6. Emotion type on each of the categories. | |
|---|---|
| **Categories** | **Emotion(anger/anticipation/ disgust/fear/joy/love/optimism/pessimism/sadness/surprise/trust)** |
| Indirect Harassment(260) | (10/18/0/0/120/0/1/0/80/0) |
| Sexual Harassment(417) | (64/0/229/0/83/0/0/0/41/0/0) |
| Physical Harassment(123) | (55/0/43/0/0/0/0/4/21/0/0) |

is in line with the previous results.

## 8. Conclusion

In this paper, we tried to focus on a subject which has attracted a lot of attention these days. We used Semeval task1: affect in tweets and tried to understand the emotion type and intensity of emotion in each category of sexual harassment. After training algorithms, we picked the algorithm with the highest accuracy and tested it on the sexual harassment tweets. This work is the first work of its type and shows there are some similarities in the physical and sexual harassment categories. Indirect harassment, known as benevolent harassment, inhabits a mild range of intensity while the other two have very high intensity of disgust, anger, sadness and even joy. It shows not only users trying to show true feeling with the intensity toward women; they enjoy sending the sexist tweets too. This is a nice avenue to follow and as future work, working on transfer learning algorithms, zero shot learning along with different character and word ngrams can be considered.